\documentclass[preprint,preprintnumbers,amsmath,amssymb]{revtex4}

\begin{document}

\title{Chirally symmetric effective field theory for nuclei}

\author{Andrzej Staszczak}%
\email{stas@tytan.umcs.lublin.pl}
\affiliation{Institute of Physics, Maria Curie-Sk{\l}odowska University,\\
pl. M. Curie-Sk{\l}odowskiej 1, 20-031 Lublin, Poland}

\date{\today}

\begin{abstract}
The Lorentz-invariant nuclear lagrangian of Furnstahl, Serot and
Tang (FST) is discussed. The FST lagrangian is derived in terms of
an effective field theory and exhibits a nonlinear realization of
chiral symmetry $SU(2)_L\times SU(2)_R$. The relevant degrees of
freedom are nucleons, pions and the low-lying non-Goldstone
bosons: isoscalar scalar ($\sigma$) and vector ($\omega$) mesons,
and isovector vector ($\rho$) mesons. The terms in the lagrangian
are organized by applying Georgi's naive dimensional analysis and
naturalness condition. As a consequence all coupling constants in
theory are dimensionless and of order unity.
\end{abstract}


\maketitle

\section{Introduction}
The effective field theory (EFT) technique allows to construct in
a controlled manner, below a characteristic energy scale, the most
general lagrangian consistent with relevant degrees of freedom and
symmetries of an underlying theory. In nuclear physics, the EFT
method relies on the symmetries of QCD to construct the effective
lagrangian. The main component of this construction is the chiral
$SU(2)_{L}\times SU(2)_{R}$ symmetry. This almost perfect symmetry
is spontaneously broken to its vectorial subgroup $SU(2)_{V=L+R}$
with the appearance of pseudo-Goldstone bosons (pions).

One of the recent attempts at formulating EFT for finite nuclei
and nuclear matter is the generalization of Walecka quantum
hadrodynamics (QHD) \cite{WAL74,SW86} proposed by Furnstahl, Serot
and Tang (FST) \cite{FST97,SW97}. The FST lagrangian is derived by
expansion in powers of the lowest lying hadronic fields and their
derivatives. The relevant degrees of freedom are nucleons, pions,
isoscalar-vector fields ($\omega$-mesons), isoscalar-scalar fields
($\sigma$-mesons), and isovector-vector fields ($\rho$-mesons). In
this lagrangian the chiral symmetry is realized nonlinearly using
a standard procedure of Weinberg \cite{WCCWZ1}, and Callen,
Coleman, Wess, Zumino \cite{WCCWZ2}, (WCCWZ). The terms of the
lagrangian are organized by applying Georgi's \cite{GEO84,GEO93}
naive dimensional analysis (NDA) and a ``naturalness'' condition.

The framework of EFTs and Georgi's naive dimensional analysis are
detailed in Sect. \ref{sec:eft2}. In Sect. \ref{sec:eft3} the nonlinear 
realization of chiral symmetry and the FST chirally symmetric effective
lagrangian are discussed. The Dirac-Hartree approximation of the
FST lagrangian, \textit{i.e.}, a treatment of the lagrangian at
the level of classical meson fields and valence nucleons is shown
in Sect. \ref{sec:eft4}. A short summary is given in the last section.

\section{Effective Field Theories}\label{sec:eft2}
\subsection{Principles of the EFTs}
An essential idea underlying the effective field theories (EFTs),
see \textit{e.g.}
Ref. \cite{DGH,WEIN,Nowak,Dobado,Polch,Geor,Ecker}, is relevant to
the appearance of disparate characteristic energy scales, $E\ll
E_0$, in quantum field theories. Suppose that we are interested in
physics at lower scale $E$, then we can choose a cut-off scale
$\Lambda$ at or slightly below $E_0$ and divide the generic fields
$\phi$ into two parts: a low- $\phi_L$ and a high-energy $\phi_H$
($\phi = \phi_L + \phi_H$), according their momenta are smaller or
greater than $\Lambda$
\begin{equation}
\phi_L(\mbox{\boldmath$k$}) \! : \
\left|\mbox{\boldmath$k$}\right| \! < \Lambda, \qquad
\phi_H(\mbox{\boldmath$k$}) \! : \
\left|\mbox{\boldmath$k$}\right| \! \geq \Lambda . \label{eq:eft1}
\end{equation}
The effective lagrangian is obtained by path integrating over the
high-energy part $\phi_H$ in the generating functional $Z$
\begin{equation}
Z = \int\!\left[ d \phi_L \right]\left[ d \phi_H \right]
e^{i\!\int\!d^4\!x \mathcal{L}(\phi_L , \phi_H )} = \int\!\left[
d\phi_L \right] e^{i\!\int\!d^4\!x
\mathcal{L}_{\mathrm{eff}}(\phi_L )} , \label{eq:eft2}
\end{equation}
where
\begin{equation}
\int\!d^4\!x \mathcal{L}_{\mathrm{eff}}(\phi_L ) = -i \ln\!
\int\!\left[ d \phi_H \right] e^{i\!\int\!d^4\!x
\mathcal{L}(\phi_L , \phi_H )} . \label{eq:eft3}
\end{equation}
This defines the procedure of eliminating the high-energy degrees
of freedom $\phi_H$, referred to as ``decimation''. The next step
is to write $\mathcal{L}_{\mathrm{eff}}$ in terms of
\textit{local} operators $O_{i}(\phi_L )$,
\begin{equation}
\mathcal{L}_{\mathrm{eff}}(\phi_L ) =
\sum_{i}^{\infty}g_{i}(\Lambda)O_{i}(\phi_L ) , \label{eq:eft4}
\end{equation}
where $g_{i}(\Lambda)$ are the coupling constants absorbing the
contribution of the integrated out high-energy degrees of freedom
$\phi_H$.

In Eq.~(\ref{eq:eft4}) the effective lagrangian is represented by
a infinity series of interactions that involve the relevant
degrees of freedom and satisfy the assumed symmetries of the
underlying high-energy theory. In order to make this procedure
useful we need some dimensional analysis. In units $\hbar\! =\!
c\! =\! 1$ the action $S_{\mathrm{eff}} = \int\!d^4\!x
\mathcal{L}_{\mathrm{eff}}(\phi_L )$ is dimensionless. If an
operator $O_i$ has dimension $\delta_i$, $\left[ O_i \right] =
\left[ m \right]^{\delta_i} \equiv \delta_i$, then $g_i$ has
dimension $\left[ g_i \right] = 4 - \delta_i$ and we can define
dimensionless coefficients $c_i = \Lambda^{\delta_{i}-4}g_i$,
which additionally are assumed to be ``natural'', \textit{i.e.} of
order $\mathcal{O}(1)$. For a process at scale $E$, we can
estimate dimensionally the magnitude of the i'th operator in the
action as
\begin{equation}
\int\!d^4\!x O_i \sim E^{\delta_{i}-4} ,
\end{equation}
so that the i'th term is of order
\begin{equation}
\int\!d^4\!x \frac{c_i}{\Lambda^{\delta_{i}-4}}\ O_i \sim c_i
\left(\frac{E}{\Lambda}\right)^{\delta_{i}-4} .
\end{equation}
Now we can see that at energies below $\Lambda$, the behaviour of
the different operators is determined by their dimension. If
$\delta_i < 4$ the operator is more and more important when $E
\rightarrow 0$, and is termed \textit{relevant}. Similarly, if
$\delta_i > 4$ the operator is less and less important, and is
termed \textit{irrelevant}. An operator with $\delta_i = 4$ is
equally important at all scales and is called \textit{marginal}.

At energies much below $\Lambda$, corrections due to the
\textit{irrelevant} (non-renormalizable) parts are suppressed by
powers of $E /\! \Lambda$ and the effective lagrangian is able to
describe the low-energy physics. The accurate procedure which
connects the order of the expansion in powers $E /\! \Lambda$ with
the terms in the effective lagrangian that need to be included at
that order is called ``power counting''.

\subsection{Naive dimensional analysis}
There are at least two relevant energy scales in nuclear physics:
the pion-decay constant $f_{\pi}\! \approx 93$ MeV and the larger
scale $\Lambda\! \sim\! 4 \pi f_{\pi}\! \sim\! 1$ GeV, which
characterizes the mass scale of physics beyond Goldstone bosons.
Using a naive dimensional analysis (NDA) proposed by Georgi and
Manohar \cite{GEO84,GEO93}, for assigning the LECs of appropriate
sizes, the effective lagrangian describing interactions of
nucleons $N(x)$, pions $\vec{\pi}(x)$, and non-Goldstone bosons
(scalar $\phi(x)$ and/or vector $V(x)$ mesons) takes a general
form
\begin{eqnarray}
\mathcal{L}_{\mathrm{eff}} &=&
\sum_{\left\{ndpb\right\}}^{\infty}\! c_{ndpb}\!
\left(\frac{\overline{N} \Gamma N}{f_{\pi}^2
\Lambda}\right)^{\frac{n}{2}}\! \left(\frac{\mathcal{D},
m_{\pi}}{\Lambda}\right)^{d}\!
\left(\frac{\vec{\pi}}{f_{\pi}}\right)^{p}\!
\frac{1}{b!}\left(\frac{\phi, V}{f_{\pi}}\right)^{b}\!
f_{\pi}^{2} \Lambda^{2} \label{eq:chi1} \\
&=& \sum_{\Delta=0}^{\infty} \mathcal{L}^{(\Delta)} ,
\label{eq:chi2}
\end{eqnarray}
where $\Gamma$ is a product of Dirac matrices, $\mathcal{D}$ a
covariant derivative, $m_{\pi}$ a pion mass (treated as
derivative) and $c_{ndpb}$ the dimensionless LECs which are
assumed to be natural, of $\mathcal{O}(1)$.

In Eq.~(\ref{eq:chi2}) the interactions are grouped in sets
$\mathcal{L}^{(\Delta)}$ of common index $\Delta \equiv
\frac{n}{2}+d+b-2$, according to (\ref{eq:chi1}), each of them
carries a factor of the form $(1/\Lambda)^{\Delta}$. This formula
has profound implications if we invoke chiral symmetry,
Ref. \cite{WEIN79}. For strong interactions (in absence of external
gauge fields, \textit{e.g.} photons) chiral constraint guarantees
that $\Delta \geq 0$ and the large scale $\Lambda\! \sim\! 4 \pi
f_{\pi}\! \sim\! 1$ GeV does not occur with positive powers in
Eq.~(\ref{eq:chi1}).

\section{Effective Chiral Lagrangian for Nuclei}\label{sec:eft3}
\subsection{Nonlinear realization of chiral symmetry}
Let us briefly collect the basic ingredients for considering the
chiral effective lagrangian. In the chiral limit, where $N_f = 2$
or $3$ of quarks are massless (\textit{u}, \textit{d} and possibly
\textit{s}), the underlying QCD lagrangian is invariant under a
global group
\begin{equation}
U(N_f)_{L}\times U(N_f)_{R}\ \simeq\ SU(N_f)_{L}\times
SU(N_f)_{R}\times U(1)_{V}\times U(1)_{A} . \label{eq:qcd1}
\end{equation}
However, at the quantum level, due to the axial anomaly, the
$U(1)_{A}$ symmetry is broken that, \textit{e.g.}, leads to
nonzero mass of $\eta'(958)$ meson even in the chiral limit. In
the hadronic world, the chiral group $G\equiv SU(N_f)_{L}\times
SU(N_f)_{R}$ is spontaneously broken in the vacuum
\begin{equation}
SU(N_f)_{L}\times SU(N_f)_{R}\times U(1)_{V}\ \longrightarrow\
SU(N_f)_{V=L+R}\times U(1)_{V} , \label{eq:qcd2}
\end{equation}
to the vectorial subgroup $H\equiv SU(N_f)_{V}$, either of isospin
when $N_f=2$ or flavor $SU(3)$ when $N_f=3$. The preserve vector
group $U(1)_{V}$ is realized as baryon number conservation.
According to Goldstone's theorem the number of Goldstone fields is
the dimension of coset space $G/H$, which is the number of
generators of $G$ that are not also generators of $H$. In our case
of chiral symmetry $\mathrm{dim}\; G/H= N^{2}_{f}-1$ and we can
identify $N^{2}_{f}-1$ pseudoscalar Goldstone bosons,
$\varphi\equiv\pi[,K,\eta]$, with the pions for $N_f=2$, plus the
kaons and an eta meson for $N_f=3$.

The nonlinear realization of spontaneously broken chiral symmetry,
denoted by WCCWZ, was suggested by Weinberg \cite{WCCWZ1} and
developed further by Callan, Coleman, Wess, and
Zumino \cite{WCCWZ2}. In the WCCWZ formalism the Goldstone bosons
$\varphi$, being coordinates of the coset space $G/H$, are
naturally represented by elements $\xi(x)=\xi(\varphi(x))$ of this
coset space. The chiral symmetry is defined by specifying the
action of $G$ on the representative $\xi(\varphi)$, with a
canonical choice of coset representative this transformation takes
the form
\begin{equation}
\xi(\varphi) \stackrel{g}{\longrightarrow}
g_R\xi(\varphi)h^{\dag}(g,\varphi) =
h(g,\varphi)\xi(\varphi)g_L^{\dag}, \label{eq:ccwz0}
\end{equation}
where $g\equiv(g_L,g_R)\in G$. The equality in (\ref{eq:ccwz0}) is
due to parity and it defines the so-called compensator (field)
$h(g,\varphi(x))\in H$. Its dependence on the Goldstone boson
fields $\varphi(x)$ is a characteristic feature of the nonlinear
realization of chiral symmetry.

Let us restrict ourselves to the two flavor case $N_f=2$ with the
isotriplet of pions collected in a $2 \times 2$ special unitary
matrix
\begin{equation}
\pi(x)\equiv\vec{\pi}(x)\cdot{\textstyle\frac{1}{2}}\vec{\tau} =
{\textstyle\frac{1}{2}}\left( \begin{array}{cc}
             \pi^0            & {\sqrt{2}} \pi^+ \\
             {\sqrt{2}} \pi^- & -\pi^0
                            \end{array}\right) , \label{eq:ccwz1}
\end{equation}
where $\vec{\tau}$ are Pauli matrices. Applying an exponential
parametrization a coset representative $\xi(x)=\xi(\pi(x))$ can be
written as
\begin{equation}
U(x)\equiv\xi^{2}(x)=\exp\left(2i\pi(x)/f_{\pi}\right) ,
\label{eq:ccwz2}
\end{equation}
where $f_{\pi}\approx 93$ MeV is the pion-decay constant. The
isospinor nucleon field is represented by a column matrix
\begin{equation}
 N(x) = \left( \begin{array}{c}%
                   p(x)\\
                   n(x)
               \end{array} \right) , \label{eq:ccwz3}
\end{equation}
where $p(x)$ and $n(x)$ are the proton and neutron fields,
respectively. The relevant low-lying non-Goldstone bosons are an
isovector-vector $\rho(770)$ meson
$\rho_{\mu}(x)\equiv\vec{\rho}_{\mu}(x)\cdot\frac{1}{2}\vec{\tau}$,
an isoscalar-vector meson $\omega(782)$ represented by a vector
field $V_{\mu}(x)$ and an effective isoscalar-scalar field $S(x)$
($\sigma$ meson) is described by the shifted field $\phi(x)\equiv
S_0-S(x)$, where $S_0$ is the vacuum expectation value of the
scalar field $S$, see Ref. \cite{FST95}. The $\omega$ meson is
needed to describe the short-range repulsion and $\phi$ effective
field is included to incorporate the mid-range attraction of the
$NN$ interaction.

Following WCCWZ, the nonlinear realization of the chiral symmetry
for the mentioned above degrees of freedom are
\begin{eqnarray}
\xi(x)          &\stackrel{g}{\longrightarrow}&
g_R\,\xi(x)\,h^{\dag}\big(g,\pi(x)\big) =
h\big(g,\pi(x)\big)\,\xi(x)\,g_L^{\dag} , \label{eq:ccwz4} \\
N(x)            &\stackrel{g}{\longrightarrow}&
h\big(g,\pi(x)\big)\,N(x),\qquad
\overline{N}(x)\stackrel{g}{\longrightarrow}
\overline{N}(x)\,h^{\dag}\big(g,\pi(x)\big) , \label{eq:ccwz5} \\
\rho_{\mu}(x)   &\stackrel{g}{\longrightarrow}&
h\big(g,\pi(x)\big)\,\rho_{\mu}(x)\,h^{\dag}\big(g,\pi(x)\big) ,
\label{eq:ccwz6}
\end{eqnarray}
where $g=(g_L,g_R)\in SU(2)_{L}\times SU(2)_{R}$. The matrix field
$U(x)$ transforms linearly under chiral transformations
\begin{equation}
U(x)\stackrel{g}{\longrightarrow}g_LU(x)g^{\dag}_R ,
\label{eq:ccwz7}
\end{equation}
and the other degrees of freedom, neutral-isoscalar fields
$V_{\mu}(x)$ and $\phi(x)$, can be treated as the chiral singlets.

We can also define an axial vector field $a_{\mu}$ and a polar
vector field $v_{\mu}$
\begin{eqnarray}
a_{\mu} &\equiv&
-{\frac{i}{2}}(\xi^{\dag}\partial_{\mu}\xi-\xi\partial_{\mu}
\xi^{\dag})=a_{\mu}^{\dag}, \label{eq:ccwz8} \\
v_{\mu} &\equiv&
-{\frac{i}{2}}(\xi^{\dag}\partial_{\mu}\xi+\xi\partial_{\mu}%
\xi^{\dag})=v_{\mu}^{\dag}\ , \label{eq:ccwz9}
\end{eqnarray}
where the hermiticity follows from $\partial_{\mu}(\xi^{\dag}\xi)
= 0 = \partial_{\mu}(\xi\xi^{\dag})$. Under the chiral symmetry
the transformation of $a_{\mu}$ is homogeneous
\begin{equation}
a_{\mu}\stackrel{g}{\longrightarrow}ha_{\mu}h^{\dag} ,
\label{eq:ccwz10}
\end{equation}
whereas that of $v_{\mu}$ is inhomogeneous
\begin{equation}
v_{\mu}\stackrel{g}{\longrightarrow}h(v_{\mu}
-i\partial_{\mu})h^{\dag} . \label{eq:ccwz11}
\end{equation}
In fact $v_{\mu}$ is the connection on the coset space and with it
we can construct the covariant derivatives on this space. For
example, since transformations (\ref{eq:ccwz5}) and
(\ref{eq:ccwz6}) are not only nonlinear but also \textit{local},
via compensator field $h(g,\pi(x))$, that requires the
introduction of the chirally covariant derivatives for nucleon and
rho meson fields:
\begin{equation}
D_{\mu}N=(\partial_{\mu}+iv_{\mu})N ,\qquad
D_{\mu}\rho_{\nu}=\partial_{\mu}\rho_{\nu}+i[v_{\mu}, \rho_{\nu}]
, \label{eq:ccwz12}
\end{equation}
which transform covariantly under the chiral group. Also, the
curvature (strength) tensor, $v_{\mu\nu}$, associated with the
connection can be expressed in terms of axial vector fields,
$a_{\mu}$, as
\begin{equation}
v_{\mu\nu}=\partial_{\mu}v_{\nu}-\partial_{\nu}v_{\mu}+i[v_{\mu},v_{\nu}]
= -i[a_{\mu },a_{\nu}] . \label{eq:ccwz13}
\end{equation}
The covariant derivative of rho meson can be used to construct the
covariant field tensor
\begin{equation}
\quad\rho_{\mu\nu}=
D_{\mu}\rho_{\nu}-D_{\nu}\rho_{\mu}+ig_{\rho}[\rho_{\mu},\rho_{\nu}]
, \label{eq:ccwz14}
\end{equation}
which is the last ingredient needed to build the FST effective
chiral lagrangian.

\subsection{FST effective chiral lagrangian}
The low-energy effective lagrangian of Furnstahl, Serot and Tang
(FST) \cite{FST97,SW97}, see also Ref. \cite{FS00,Fur03,Ser03},
incorporates the symmetries of QCD: Lorentz invariance, parity
invariance, nonlinear realization of chiral $SU(2)_{L}\times
SU(2)_{R}$ symmetry, this lagrangian is also invariant under the
electromagnetic $U(1)_{em}$ and isospin $SU(2)$ groups. The FST
lagrangian is expanded in powers of fields and their derivatives
in the procedure of power counting with index $\tilde{\Delta}
=\frac{n}{2}+d+b$, where $n$ is the number of nucleon fields, $d$
is the number of derivatives and $b$ is the number of
non-Goldstone boson fields in each term. Taking as the large
energy scale $\Lambda$ in Eq. (\ref{eq:chi1}) the nucleon mass
$M\!=\!939$ MeV, we may write the effective chiral lagrangian
through quartic order ($\tilde{\Delta}\leq 4$) as the sum
\begin{equation}
\mathcal{L}_{\mathrm{eff}}(x)=
\mathcal{L}^{\scriptscriptstyle{(4)}}_{\mathrm{N}}(x)+
\mathcal{L}^{\scriptscriptstyle{(4)}}_{\mathrm{M}}(x)+
\mathcal{L}^{\scriptscriptstyle{(4)}}_{\mathrm{EM}}(x)\ .
\label{eq:Lfull}
\end{equation}
The part of lagrangian involving nucleons takes the form
\begin{eqnarray}
\mathcal{L}^{\scriptscriptstyle{(4)}}_{\mathrm{N}}(x) &=&
\overline{N}\Big[\gamma^{\mu}\left(i\partial_{\mu}-v_{\mu}-
g_{\rho}\rho_{\mu}-g_{\mathrm{v}}V_{\mu}\right)+
g_{\mathrm{\scriptscriptstyle A}}\gamma^{\mu}\gamma _{5}a_{\mu}
-\left(M-g_{\mathrm{s}}\phi\right)\Big]N  \nonumber \\ [2pt]
                                  &&
-{\frac{f_{\rho }g_{\rho }}{4M}}\overline{N}\rho _{\mu\nu}
\sigma^{\mu\nu}N-{\frac{f_{\mathrm{v}}g_{\mathrm{v}}}{4M}}
\overline{N}V_{\mu\nu}\sigma^{\mu\nu}N-{\frac{\kappa _{\pi}}{M}}
\overline{N}v_{\mu\nu}\sigma^{\mu\nu}N  \nonumber \\ [2pt]
                                   &&
+\cdots , \label{eq:Ln}
\end{eqnarray}
where
$\sigma_{\mu\nu}\!=\!\frac{i}{2}[\gamma_{\mu},\gamma_{\nu}]$,
$V_{\mu\nu}\equiv \partial_{\mu}V_{\nu}-\partial_{\nu}V_{\mu}$ is
the covariant tensor of the $\omega$ meson,
$g_{\mathrm{\scriptscriptstyle A}}\!\approx\!1.26$ is the axial
coupling constant, $g_{\rho},\,f_{\rho}$, and
$g_{\mathrm{v}},\,f_{\mathrm{v}}$ are vector and so-called tensor
couplings for $\rho$ and $\omega$ mesons, see Ref.\cite{Mach89},
$g_{\mathrm{s}}$ is a Yukawa coupling for the effective scalar
field $\phi$, and $\kappa_{\pi}\!=\!\frac{f_{\rho}}{4}$ is the
coupling for higher-order $\pi N$ interaction. The ellipsis
represents redundant or tiny additional terms with $\pi N$ and
$\pi \pi$ interactions, which have been omitted in the FST
lagrangian.

The mesonic part of the effective lagrangian is
\begin{eqnarray}
\mathcal{L}^{\scriptscriptstyle{(4)}}_{\mathrm{M}}(x) &=&
{\textstyle\frac{1}{2}}\Big(1+\alpha_{1}\frac{g_{\mathrm{s}}
\phi}{M}\Big)\partial_{\mu}\phi\partial^{\mu}\phi+\frac{f_{\pi}^2}{4}\,
\mathrm{tr}\,(\partial_{\mu}U\partial^{\mu}U^{\dagger})
\nonumber \\[2pt]
                                  &&
-{\textstyle\frac{1}{2}}\,\mathrm{tr}\,(\rho_{\mu\nu}\rho^{\mu\nu})-
{\textstyle\frac{1}{4}}\Big(1+\alpha_{2}\frac{g_{\mathrm{s}}\phi}
{M}\Big)V_{\mu\nu}V^{\mu\nu}-g_{\rho\pi\pi}\frac{2f_{\pi}^2}{m_{\rho}^2}
\,\mathrm{tr}\,(\rho_{\mu\nu}v^{\mu\nu})  \nonumber \\[2pt]
                                  &&
+{\textstyle\frac{1}{2}}\Big(1+\eta_1\frac{g_{\mathrm{s}}\phi}{M}+
\frac{\eta_2}{2}\frac{g_{\mathrm{s}}^2\phi^2}{M^2}\Big)
m_{\mathrm{v}}^2V_{\mu}V^{\mu}+\frac{1}{4!}\zeta_{0}
g_{\mathrm{v}}^2(V_{\mu}V^{\mu})^2   \label{eq:Lm} \\[2pt]
                                  &&
+\Big(1+\eta_{\rho}\frac{g_{\mathrm{s}}\phi}{M}\Big)m_{\rho}^2
\,\mathrm{tr}\,(\rho_{\mu}\rho^{\mu})-m_{\mathrm{s}}^2\phi^2
\Big({\textstyle\frac{1}{2}}+\frac{\kappa_3}{3!}\,
\frac{g_{\mathrm{s}}\phi}{M}+\frac{\kappa_4}{4!}\,
\frac{g_{\mathrm{s}}^2\phi^2}{M^2}\Big) , \nonumber
\end{eqnarray}
where $m_{\mathrm{v}}=782$ MeV, $m_{\rho}=770$ MeV, and
$m_{\mathrm{s}}$ are $\omega$, $\rho$ and $\sigma$ mesons masses,
$g_{\rho\pi\pi}$ is the coupling of $\rho\pi\pi$ interaction,
which (assuming vector-meson dominance) is
$g_{\rho\pi\pi}=g_{\rho}$. The trace ``$\mathrm{tr}$'' is in the
$2\times 2$ isospin space.

The electromagnetic interactions are described by
\begin{eqnarray}
\mathcal{L}^{\scriptscriptstyle{(4)}}_{\mathrm{EM}}(x) &=&
-{\textstyle\frac{1}{4}}F^{\mu\upsilon}F_{\mu\nu}-
{\textstyle\frac{1}{2}}e\overline{N}\gamma^{\mu}(1+\tau_3)NA_{\mu}
-\frac{e}{4M}F_{\mu\nu}\overline{N}\lambda\sigma^{\mu\nu}N
\nonumber \\[2pt]                                   &&
-\frac{e}{2M^2}\overline{N}\gamma_{\mu}(\beta_{\mathrm{s}}+
\beta_{\mathrm{v}}\tau_{3})N\partial_{\nu}F^{\mu\nu}
-2ef_{\pi}^2A^{\mu}\,\mathrm{tr}\,(v_{\mu}\tau_{3})
\nonumber \\[2pt]                                   &&
-\frac{e}{2g_{\gamma}}F_{\mu\nu}\Big[\,\mathrm{tr}\,
(\tau_3\rho^{\mu\nu})+{\textstyle\frac{1}{3}}V^{\mu\nu}\Big]+\cdots
, \label{eq:Lem}
\end{eqnarray}
where $A_{\mu}$ is the electromagnetic field, $F_{\mu\nu}$ is the
electromagnetic field-strength tensor. According to vector-meson
dominance and phenomenology one can find that $g_{\gamma}=5.01$.
The lagrangian
$\mathcal{L}^{\scriptscriptstyle{(4)}}_{\mathrm{EM}}$ is invariant
under the $U(1)_{em}$ group. The composite structure of the
nucleon is included through an anomalous moment $\lambda\equiv
\frac{1}{2}\lambda_{\mathrm{p}}(1+\tau_{3})+
\frac{1}{2}\lambda_{\mathrm{n}}(1-\tau_{3})$, with
$\lambda_{\mathrm{p}}=1.793$ and $\lambda_{\mathrm{n}}=-1.913$ the
anomalous magnetic moments of the proton and the neutron,
respectively. The ellipsis represents redundant terms of
$\mathcal{O}(e^2)$.

The effective chiral lagrangian Eq. (\ref{eq:Lfull}) at a given
order contains certain parameters that are not constrained by the
symmetries, the so-called low-energy constants (LECs). Apart from
the isoscalar ($\beta_{\mathrm{s}}$), isovector
($\beta_{\mathrm{v}}$) electromagnetic form factors and the tensor
coupling for $\rho$ meson ($f_{\rho}$), which are fixed from the
free-space charge radii of the nucleon, the remaining thirteen
LECs
$\{\,\frac{g_{\mathrm{s}}}{4\pi},\,\frac{g_{\mathrm{v}}}{4\pi},\,
\frac{g_{\rho}}{4\pi},\,\eta_{1},\,\eta_{2},\,\eta_{\rho},\,
\kappa_{3},\,\kappa_{4},\,\zeta_{0},\,\frac{m_{\mathrm{s}}}{M},\,
f_{\mathrm{v}},\,\alpha_{1},\,\alpha_{2}\}$ have to be determined
from experimental data. The LECs are defined applying the naive
dimensional analysis so that they are assumed to be of order
unity, \textit{i.e.} ``natural''.

\section{Dirac-Hartree Approximation}\label{sec:eft4}
The mean-field approximation (ignores) dismisses all quantum
fluctuations of the meson fields and treats them by their
expectation values. Assuming the time reversal invariance the
spatial components of the vector field vanish and we can define
scaled mean meson fields (potentials) by including couplings:
$W(\mbox{\boldmath$r$})=g_{\mathrm{v}}V_{0}(\mbox{\boldmath$r$})$,
$\mathit{\Phi}(\mbox{\boldmath$r$})=g_{\mathrm{s}}\phi_{0}
(\mbox{\boldmath$r$})$,
$R(\mbox{\boldmath$r$})=g_{\rho}b_{0}(\mbox{\boldmath$r$})$ and
$A(\mbox{\boldmath$r$})=eA_{0}(\mbox{\boldmath$r$})$. Since the
nuclear ground state has a well-defined charge, only the neutral
rho meson field (denoted by $b_0$) has been used, also since the
ground state is assumed to have well-defined parity the
pseudo-scalar pion field does not contribute in this
approximation.

If we restrict consideration to static nuclear systems the Dirac
equation with eigenvalues $E_{\alpha}$ and eigenfunctions
$\psi_{\alpha}(\mbox{\boldmath$r$})$ is, see Ref.\cite{SW86},
\begin{equation}
h\psi_{\alpha}(\mbox{\boldmath$r$})=
E_{\alpha}\psi_{\alpha}(\mbox{\boldmath$r$}) , \quad \int
d^3x\psi^{\dag}_{\alpha}(\mbox{\boldmath$r$})
\psi_{\alpha}(\mbox{\boldmath$r$})=1 , \label{eq:mf0}
\end{equation}
with
\begin{eqnarray}
h(\mbox{\boldmath$r$}) &=&
-i\mbox{\boldmath$\alpha\cdot\nabla$}+W(\mbox{\boldmath$r$})+
{\textstyle\frac{1}{2}}\tau_{3}R(\mbox{\boldmath$r$})+
\beta\Big[M-\mathit{\Phi}(\mbox{\boldmath$r$})\Big]+
{\textstyle\frac{1}{2}}(1+\tau_{3})A(\mbox{\boldmath$r$}) \nonumber \\[2pt]
                       &&
-\frac{i}{2M}\beta\,\mbox{\boldmath$\alpha\cdot$}
\Big[f_{\rho}{\textstyle\frac{1}{2}}\tau_{3}\mbox{\boldmath$\nabla$}
R(\mbox{\boldmath$r$})+
f_{\mathrm{v}}\mbox{\boldmath$\nabla$}W(\mbox{\boldmath$r$})\Big]+
\frac{1}{2M^2}\left(\beta_{\mathrm{s}}+\beta_{\mathrm{v}}\tau_{3}
\right)\Delta A(\mbox{\boldmath$r$}) \nonumber \\[2pt]
                       &&
-\frac{i}{2M}\lambda\beta\,\mbox{\boldmath$\alpha\cdot\nabla$}
A(\mbox{\boldmath$r$}) , \label{eq:mf1}
\end{eqnarray}
where $\beta=\gamma_0$,
$\mbox{\boldmath$\alpha$}=\gamma_0\mbox{\boldmath$\gamma$}$.

The mean-field equations for $\mathit{\Phi}(\mbox{\boldmath$r$})$,
$W(\mbox{\boldmath$r$})$, $R(\mbox{\boldmath$r$})$ and
$A(\mbox{\boldmath$r$})$ are
\begin{eqnarray}
\left(-\Delta +
m^2_{\mathrm{s}}\right)\mathit{\Phi}(\mbox{\boldmath$r$}) &=&
g^2_{\mathrm{s}}\rho_{\mathrm{s}}(\mbox{\boldmath$r$})
-\frac{m^2_{\mathrm{s}}}{M}\mathit{\Phi}^2(\mbox{\boldmath$r$})
\Big[\frac{\kappa_3}{2}+\frac{\kappa_4}{3!}
\frac{\mathit{\Phi}(\mbox{\boldmath$r$})}{M}\Big] \nonumber \\[2pt]
&& +\frac{g^2_{\mathrm{s}}}{2M}\Big[\eta_1+\eta_2
\frac{\mathit{\Phi}(\mbox{\boldmath$r$})}{M}\Big]
\frac{m^2_{\mathrm{v}}}{g^2_{\mathrm{v}}}W^2(\mbox{\boldmath$r$})+
\frac{g^2_{\mathrm{s}}\eta_{\rho}}{2M}
\frac{m^2_{\rho}}{g^2_{\rho}}R^2(\mbox{\boldmath$r$}) \nonumber \\
[2pt] && +\frac{\alpha_1}{2M}\Big[\big(\mbox{\boldmath$\nabla$}
\mathit{\Phi}(\mbox{\boldmath$r$})\big)^2+
2\mathit{\Phi}(\mbox{\boldmath$r$})\Delta
\mathit{\Phi}(\mbox{\boldmath$r$})\Big]  \nonumber \\
&&+\frac{\alpha_2g^2_{\mathrm{s}}}{2Mg^2_{\mathrm{v}}}
\big(\mbox{\boldmath$\nabla$}W(\mbox{\boldmath$r$})\big)^2 ,
\label{eq:mf2}
\end{eqnarray}
\begin{eqnarray}
\left(-\Delta + m^2_{\mathrm{v}}\right)W(\mbox{\boldmath$r$}) &=&
g^2_{\mathrm{v}}\Big[\rho_{\mathrm{B}}(\mbox{\boldmath$r$})+
\frac{f_{\mathrm{v}}}{2M}\mbox{\boldmath$\nabla\cdot$}
\big(\widehat{\mbox{\boldmath$r$}}\rho^{\mathrm{T}}_{\mathrm{B}}
(\mbox{\boldmath$r$})\big)\Big] \nonumber \\[2pt]
&& -\Big[\eta_1+\frac{\eta_2}{2}
\frac{\mathit{\Phi}(\mbox{\boldmath$r$})}{M}\Big]
\frac{\mathit{\Phi}(\mbox{\boldmath$r$})}{M}m^2_{\mathrm{v}}
W(\mbox{\boldmath$r$})-\frac{1}{3!}\zeta_{0}
W^3(\mbox{\boldmath$r$}) \nonumber \\[2pt]
&& +\frac{\alpha_2}{M}\Big[\mbox{\boldmath$\nabla$}
\mathit{\Phi}(\mbox{\boldmath$r$})\cdot
\mbox{\boldmath$\nabla$}W(\mbox{\boldmath$r$})+
\mathit{\Phi}(\mbox{\boldmath$r$})\Delta
W(\mbox{\boldmath$r$})\Big] \nonumber \\[2pt]
&& -\frac{e^{2}g_{\mathrm{v}}}{3g_{\gamma}}\,\rho_{\mathrm{chg}}
(\mbox{\boldmath$r$}) , \label{eq:mf3}
\end{eqnarray}
\begin{eqnarray}
\left(-\Delta + m^2_{\rho}\right)R(\mbox{\boldmath$r$}) &=&
{\textstyle\frac{1}{2}}g^2_{\rho}\Big[\rho_{3}(\mbox{\boldmath$r$})+
\frac{f_{\rho}}{2M}\mbox{\boldmath$\nabla\cdot$}
\big(\widehat{\mbox{\boldmath$r$}}\rho^{\mathrm{T}}_{3}
(\mbox{\boldmath$r$})\big)\Big] \nonumber \\[2pt]
&& -\eta_{\rho}\frac{\mathit{\Phi}(\mbox{\boldmath$r$})}{M}
m^2_{\rho}R(\mbox{\boldmath$r$})-\frac{e^{2}g_{\rho}}
{g_{\gamma}}\,\rho_{\mathrm{chg}}(\mbox{\boldmath$r$}) ,
\label{eq:mf4}
\end{eqnarray}
\begin{equation}
-\Delta A(\mbox{\boldmath$r$})=
e^2\rho_{\mathrm{chg}}(\mbox{\boldmath$r$}) . \label{eq:mf5}
\end{equation}

Assuming spherical symmetry and parity conservation the
eigenfunctions of Dirac equation (\ref{eq:mf0}) (the
positive-energy spinors) can be written as
\begin{equation}
\psi_{\alpha}(\mbox{\boldmath$r$})\equiv\psi_{n\kappa mt}
(\mbox{\boldmath$r$})= \left( \begin{array}{c}
     i\left[G_{a}(r)/r\right]\mathit{\Phi}_{\kappa m} \\
     -\left[F_{a}(r)/r\right]\mathit{\Phi}_{-\kappa m}
                              \end{array}\right)\zeta_{t} , \qquad
a\equiv\{n,\kappa,t\} , \label{eq:mf6}
\end{equation}
\begin{equation}
\int_0^\infty
dr\left(\left|G_{a}(r)\right|^2+\left|F_{a}(r)\right|^2\right)=1
\end{equation}
where $\mathit{\Phi}_{\kappa m}=\sum_{m_{\ell}m_s}\langle\ell
m_{\ell}\frac{1}{2}m_{s}|jm\rangle Y_{\ell,m_{\ell}}(\Omega)
\chi_{m_{s}}$ are spin spherical harmonics, $n$ is the principal
quantum number, $\kappa$ is a nonzero integer uniquely determining
$j$ and $\ell$ through $\kappa=(2j+1)(\ell-j)$ and $\zeta_{t}$ is
a two-component isospinor labeled by the isospin projection
$t=\frac{1}{2}$ for protons and $t=-\frac{1}{2}$ for neutrons. The
radial equations for upper ($G$) and lower ($F$) components become
\begin{eqnarray}
\Big(\frac{d}{dr}+\frac{\kappa}{r}\Big)G_{a}(r)-
\big[E_{a}-U_{1}(r)+U_{2}(r)\big]F_{a}(r)-U_{3}(r)G_{a}(r) &=& 0 ,
\label{eq:mf7} \\
\Big(\frac{d}{dr}-\frac{\kappa}{r}\Big)F_{a}(r)+
\big[E_{a}-U_{1}(r)-U_{2}(r)\big]G_{a}(r)+U_{3}(r)F_{a}(r) &=& 0 ,
\label{eq:mf8}
\end{eqnarray}
where single-particle potentials are given by
\begin{eqnarray}
U_{1}(r) &\equiv& W(r)+t_{a}R(r)+\big(t_{a}+{\textstyle
\frac{1}{2}}\big)A(r)+\frac{1}{2M^2}(\beta_{\mathrm{s}}+
2t_{a}\beta_{\mathrm{v}})\Delta A(r) , \label{eq:mf9} \\
U_{2}(r) &\equiv& M-\mathit{\Phi}(r) , \label{eq:mf10} \\
U_{3}(r) &\equiv&
\frac{1}{2M}\Big\{f_{\mathrm{v}}\frac{dW(r)}{dr}+ t_{a}f_{\rho}
\frac{dR(r)}{dr} \nonumber \\
&&+\frac{dA(r)}{dr}\big[{\textstyle \frac{1}{2}}
(\lambda_{p}+\lambda_{n})+t_{a}(\lambda_{p}-\lambda_{n})\big]\Big\}
. \label{eq:mf11}
\end{eqnarray}
The various densities that appear on the r.h.s. of the meson
equations for spherically symmetric systems are defined as
follows:
\begin{eqnarray}
\rho_{\mathrm{s}}(\mbox{\boldmath$r$}) &=&
\sum_{\alpha}^{occ}\overline{\psi}_{\alpha}(\mbox{\boldmath$r$})
\psi_{\alpha}(\mbox{\boldmath$r$})\equiv\sum_{a}^{occ}\frac{2j_a+1}
{4\pi r^2}\left(G^2_a(r)-F^2_a(r)\right) , \label{eq:mf12} \\
\rho_{\mathrm{B}}(\mbox{\boldmath$r$}) &=&
\sum_{\alpha}^{occ}\psi^{\dag}_{\alpha}(\mbox{\boldmath$r$})
\psi_{\alpha}(\mbox{\boldmath$r$})\equiv\sum_{a}^{occ}\frac{2j_a+1}
{4\pi r^2}\left(G^2_a(r)+F^2_a(r)\right) , \label{eq:mf13} \\
\rho^{\mathrm{T}}_{\mathrm{B}}(\mbox{\boldmath$r$}) &=&
\sum_{\alpha}^{occ}\psi^{\dag}_{\alpha}(\mbox{\boldmath$r$})
i\beta\mbox{\boldmath$\alpha\cdot$}\widehat{\mbox{\boldmath$r$}}
\psi_{\alpha}(\mbox{\boldmath$r$})\equiv\sum_{a}^{occ}\frac{2j_a+1}
{4\pi r^2}\,2\,G_a(r)F_a(r) , \label{eq:mf14} \\
\rho_3(\mbox{\boldmath$r$}) &=&
\sum_{\alpha}^{occ}\psi^{\dag}_{\alpha}(\mbox{\boldmath$r$})
\tau_3\psi_{\alpha}(\mbox{\boldmath$r$})\equiv\sum_{a}^{occ}
\frac{2j_a+1}{4\pi r^2}(2t_a)\left(G^2_a(r)+F^2_a(r)\right) ,
\label{eq:mf15} \\
\rho^{\mathrm{T}}_3(\mbox{\boldmath$r$}) &=&
\sum_{\alpha}^{occ}\psi^{\dag}_{\alpha}(\mbox{\boldmath$r$})
i\tau_3\beta\mbox{\boldmath$\alpha\cdot$}\widehat{\mbox{\boldmath$r$}}
\psi_{\alpha}(\mbox{\boldmath$r$})\equiv\sum_{a}^{occ}
\frac{2j_a+1}{4\pi r^2}(2t_a)\,2\,G_a(r)F_a(r)  , \label{eq:mf16}
\end{eqnarray}
where the summation superscript ``$occ$'' means that the sum runs
only over occupied (valence) orbitals up to some value of $n$ and
$\kappa$. The quantum numbers are denoted by
$\{\alpha\}=\{a;m\}\equiv\{n,\kappa,t;m\}$.

The charge density is given by
\begin{equation}
\rho_{\mathrm{chg}}(\mbox{\boldmath$r$})\equiv
\rho_{\mathrm{d}}(\mbox{\boldmath$r$})+
\rho_{\mathrm{m}}(\mbox{\boldmath$r$}), \label{eq:mf17}
\end{equation}
where the ``direct'' nucleon charge density is
\begin{equation}
\rho_{\mathrm{d}}(\mbox{\boldmath$r$})=
\rho_{\mathrm{p}}(\mbox{\boldmath$r$})+\frac{1}{2M}
\mbox{\boldmath$\nabla\cdot$}\left[\widehat{\mbox{\boldmath$r$}}
\rho_a^{\mathrm{T}}(\mbox{\boldmath$r$})\right]+
\frac{1}{2M^2}\left[\beta_{\mathrm{s}}\,\Delta
\rho_{\mathrm{B}}(\mbox{\boldmath$r$})+ \beta_{\mathrm{v}}\,\Delta
\rho_{3}(\mbox{\boldmath$r$})\right] , \label{eq:mf18}
\end{equation}
and the vector mesons contribution is
\begin{equation}
\rho_{\mathrm{m}}(\mbox{\boldmath$r$})=\frac{1}{g_{\gamma}g_{\rho}}
\Delta R(\mbox{\boldmath$r$})+\frac{1}{3g_{\gamma}g_v} \Delta
W(\mbox{\boldmath$r$}) . \label{eq:mf19}
\end{equation}
Here the ``point'' proton density and nucleon tensor density are
given by
\begin{eqnarray}
\rho_{\mathrm{p}}(\mbox{\boldmath$r$})     &\equiv&
{\textstyle\frac{1}{2}}\sum_{\alpha}^{occ}\psi^{\dag}_{\alpha}
(\mbox{\boldmath$r$})(1+\tau_3)\psi_{\alpha}(\mbox{\boldmath$r$})=
{\textstyle\frac{1}{2}}\left[\rho_{\mathrm{B}}(\mbox{\boldmath$r$})+
\rho_3(\mbox{\boldmath$r$})\right] , \label{eq:mf20}\\
\rho^{\mathrm{T}}_{a}(\mbox{\boldmath$r$}) &\equiv&
\sum_{\alpha}^{occ}\psi^{\dag}_{\alpha}(\mbox{\boldmath$r$})
i\lambda\beta\mbox{\boldmath$\alpha\cdot$}\widehat{\mbox{\boldmath$r$}}
\psi_{\alpha}(\mbox{\boldmath$r$}) , \label{eq:mf21}
\end{eqnarray}
respectively, where $\lambda$ is the anomalous magnetic moment.
Thus the spherical nuclear ground state with the presence of time
reversal symmetry is described by coupled, one-dimensional
differential equations that may be solved by an iterative
procedure. Once the solution has been found, the total energy of
the system is given by
\begin{equation}
E=\sum_a^{occ}E_a(2j_a+1)-\int d^3x\,U_m(r) , \label{eq:mf22}
\end{equation}
where
\begin{eqnarray}
U_m &\equiv&
-{\textstyle\frac{1}{2}}\mathit{\Phi}\rho_{\mathrm{s}}+
{\textstyle\frac{1}{2}}W\Big[\rho_{\mathrm{B}}+\frac{f_{\mathrm{v}}}{2M}
\mbox{\boldmath$\nabla\cdot$}\left(\widehat{\mbox{\boldmath$r$}}
\rho_{\mathrm{B}}^{\mathrm{T}}\right)\Big]+
{\textstyle\frac{1}{4}}R\Big[\rho_{3}+\frac{f_{\rho}}{2M}
\mbox{\boldmath$\nabla\cdot$}\left(\widehat{\mbox{\boldmath$r$}}
\rho_{3}^{\mathrm{T}}\right)\Big] \nonumber \\[2pt]
&& +{\textstyle\frac{1}{2}}A\rho_{\mathrm{d}}+
\frac{m^2_{\mathrm{s}}}{g^2_{\mathrm{s}}}
\frac{\mathit{\Phi}^3}{M}\Big[\frac{\kappa_3}{12}+
\frac{\kappa_4}{24}\frac{\mathit{\Phi}}{M}\Big]-
\frac{\eta_{\rho}}{4}\frac{\mathit{\Phi}}{M}
\frac{m^2_{\rho}}{g^2_{\rho}}R^2 \nonumber \\[2pt]
&& -\frac{\mathit{\Phi}}{4M}
\Big[\eta_1+\eta_2\frac{\mathit{\Phi}}{M}\Big]
\frac{m^2_{\mathrm{v}}}{g^2_{\mathrm{v}}}W^2-
\frac{1}{4!g^2_{\mathrm{v}}}\zeta_{0}W^4
+\frac{\alpha_1}{4g^2_{\mathrm{s}}}\frac{\mathit{\Phi}}{M}
\left(\mbox{\boldmath$\nabla$}\mathit{\Phi}\right)^2 \nonumber \\[2pt]
&& -\frac{\alpha_2}{4g^2_{\mathrm{v}}}\frac{\mathit{\Phi}}{M}
\left(\mbox{\boldmath$\nabla$}W\right)^2 . \label{eq:mf23}
\end{eqnarray}
One of the most prominent observables, the binding energy of a
system of $A=Z+N$ nucleons is defined by
\begin{equation}
E_B=E-E_{\mathrm{CM}}-AM , \label{eq:mf24}
\end{equation}
where $E_{\mathrm{CM}}$ is the center-of-mass (c.m.) correction
which can be estimated nonrelativistically, \textit{e.g.}, its an
empirical estimate given by Reinhard\cite{Rein89} is
$E_{\mathrm{CM}}\approx 17.2A^{-0.2}$ MeV . An older estimate from
the harmonic oscillator shell model is $E_{\mathrm{CM}}\approx
\frac{3}{4}41A^{-1/3}$ MeV.

The mean-square radius of the charge distribution, with the (c.m.)
motion correction, is given by
\begin{equation}
\langle r^2\rangle_{\mathrm{chg}}=\langle r^2\rangle
-\frac{3}{4\langle\hat{P}^2_{\mathrm{CM}}\rangle} ,
\label{eq:mf25}
\end{equation}
where
\begin{equation}
\langle r^2\rangle=\frac{1}{Z}\int d^3x\,\mbox{\boldmath$r$}^2
\rho_{\mathrm{chg}}(\mbox{\boldmath$r$}) ,\qquad
\langle\hat{P}^2_{\mathrm{CM}}\rangle=2AME_{\mathrm{CM}} .
\label{eq:mf26}
\end{equation}
Since the additional nonrenormalizable interaction between the
nucleon and electromagnetic field were included in
$\mathcal{L}_{\mathrm{EM}}$, Eq. (\ref{eq:Lem}), the charge
density $\rho_{\mathrm{chg}}$ automatically contains the effects
of nucleon structure, and it is unnecessary to introduce an
\textit{ad hoc} form factor in formula (\ref{eq:mf25}).

\section{Summary}\label{sec:eft5}
One of the major challenges in nuclear physics is to establish a
connection between nuclear dynamics and the fundamental QCD. The
chiral effective field theories are considered to offer a natural
and useful framework for this purpose. Thanks to the
implementation of nonlinear realization of chiral symmetry,
Georgi's naive dimensional analysis and the ``naturalness''
condition, the FST approach is the extension of Walecka's
hadrodynamics and may be used in nuclear physics
to \textit{cross the border} from QCD to a nuclear theory.

\end{document}